\documentclass[letterpaper]{article} % DO NOT CHANGE THIS
\usepackage[draft]{preprintlayout}  % DO NOT CHANGE THIS
\usepackage{times}  % DO NOT CHANGE THIS
\usepackage{helvet}  % DO NOT CHANGE THIS
\usepackage{courier}  % DO NOT CHANGE THIS
\usepackage[hyphens]{url}  % DO NOT CHANGE THIS
\usepackage{graphicx} % DO NOT CHANGE THIS
\usepackage{natbib}  % DO NOT CHANGE THIS AND DO NOT ADD ANY OPTIONS TO IT
\usepackage{caption} % DO NOT CHANGE THIS AND DO NOT ADD ANY OPTIONS TO IT
\usepackage[utf8]{inputenc}
\usepackage{listings}
\usepackage{algorithm}
\usepackage{algorithmic}

\usepackage{array}
\usepackage{booktabs}
\usepackage{xspace}
\usepackage[breaklinks=true]{hyperref}

\newcommand{\facet}{FACET\xspace}

\usepackage{newfloat}
\usepackage{listings}
\DeclareCaptionStyle{ruled}{labelfont=normalfont,labelsep=colon,strut=off} % DO NOT CHANGE THIS
\floatstyle{ruled}
\newfloat{listing}{tb}{lst}{}
\floatname{listing}{Listing}
\usepackage{comment}
\usepackage{xcolor}
\usepackage{listings}
\definecolor{quotebg}{RGB}{245,245,250}
\definecolor{quoteborder}{RGB}{110,120,180}
\lstdefinestyle{feedbackprompt}{
  basicstyle=\footnotesize\ttfamily,
  backgroundcolor=\color{quotebg},
  frame=single,
  framesep=8pt,
  framerule=1pt,
  rulecolor=\color{quoteborder},
  breaklines=true,
  breakatwhitespace=false,
  captionpos=b,
  showstringspaces=false,
  numbers=none,
  xleftmargin=3pt,
  xrightmargin=3pt,
  aboveskip=5pt,
  belowskip=5pt
}

\title{\facet: Teacher-Centred LLM-Based Multi-Agent Systems---\\Towards Personalized Educational Worksheets}

\author {
    Jana Gonnermann-M\"uller\textsuperscript{\rm 1,2},
    Jennifer Haase,\textsuperscript{\rm 2,3},
    Konstantin Fackeldey, \textsuperscript{\rm 4,5},
    Sebastian Pokutta\textsuperscript{\rm 4,5}
}
\affiliations {
    \textsuperscript{\rm 1}Potsdam University, Potsdam, Germany\\
    \textsuperscript{\rm 2}Weizenbaum Institute, Berlin, Germany\\
    \textsuperscript{\rm 3}Humboldt University, Berlin, Germany\\
    \textsuperscript{\rm 4}Technical University Berlin, Berlin, Germany\\
    \textsuperscript{\rm 5}Zuse Institute Berlin, Berlin, Germany\\
    Jana.Gonnermann@wi.uni-potsdam.de, Jennifer.Haase@hu-berlin.de, fackeldey@math.tu-berlin.de, Pokutta@zib.de
}
\begin{document}
\maketitle

\begin{abstract}
The increasing heterogeneity of student populations poses significant challenges for teachers, particularly in mathematics education, where cognitive, motivational, and emotional differences strongly influence learning outcomes. While AI-driven personalization tools have emerged, most remain performance-focused, offering limited support for teachers and neglecting broader pedagogical needs. This paper presents the \textit{\facet framework}, a teacher-facing, large language model (LLM)-based multi-agent system designed to generate individualized classroom materials that integrate both cognitive and motivational dimensions of learner profiles. The framework comprises three specialized agents: (1) learner agents that simulate diverse profiles incorporating topic proficiency and intrinsic motivation, (2) a teacher agent that adapts instructional content according to didactical principles, and (3) an evaluator agent that provides automated quality assurance. We tested the system using authentic grade 8 mathematics curriculum content and evaluated its feasibility through a) automated agent-based assessment of output quality and b) exploratory feedback from K-12 in-service teachers. Results from ten internal evaluations highlighted high stability and alignment between generated materials and learner profiles, and teacher feedback particularly highlighted structure and suitability of tasks. The findings demonstrate the potential of multi-agent LLM architectures to provide scalable, context-aware personalization in heterogeneous classroom settings, and outline directions for extending the framework to richer learner profiles and real-world classroom trials.
\end{abstract}

\section{Introduction} 
\label{sec:intro}

\begin{figure*}[ht]
    \centering
    \includegraphics[width=1\linewidth]{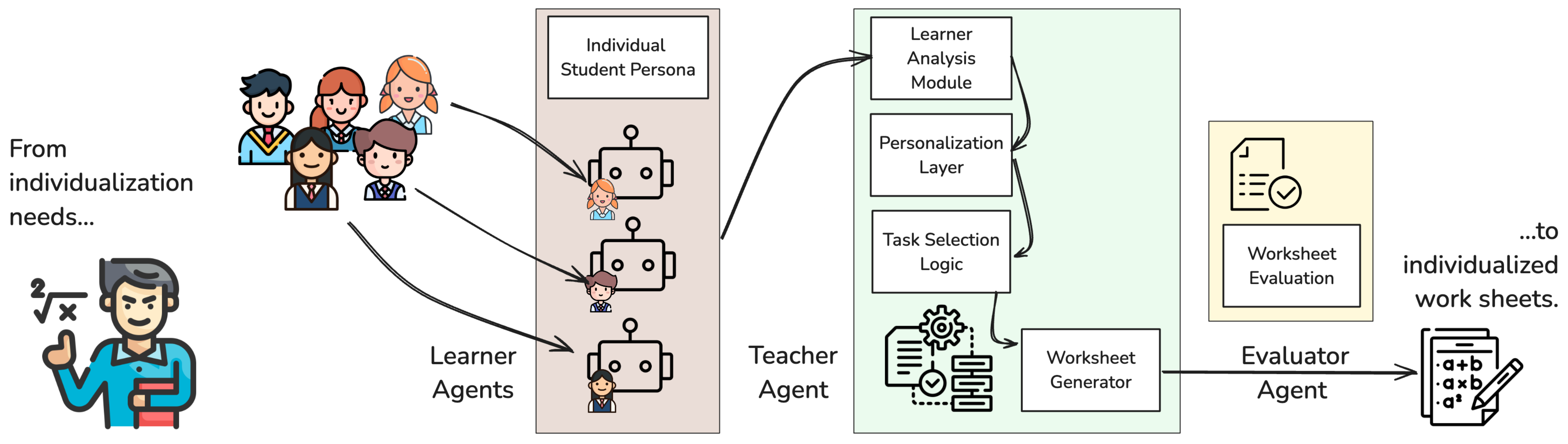}
    \caption{Core Principle for individualized classroom material using a multiple agent framework}
    \label{fig:banner}
\end{figure*}

One major contributor to the rising workload of teachers is the increasing heterogeneity of student populations. In many classrooms, students with vastly different backgrounds, needs, and skills, ranging from high achievers to those requiring significant support, are taught together. 
\citet{tomlinsonDifferentiatedInstruction2017} argues that differentiation is necessary to ensure that all students, regardless of their background, interests, or skills, have the best opportunities to learn.
Mathematics is particularly affected by these disparities, with students frequently reporting boredom, anxiety, and disengagement \citep{andersPISA2022Sind2025}. In addition, domain-specific emotional challenges are pronounced, with approximately 20\% of girls and 10\% of boys reporting anxiety in mathematics instruction, alongside similar patterns for feelings of distress \citep{andersPISA2022Sind2025}. These differences in emotional and motivational states underscore the need for personalized approaches that address both cognitive and motivational dimensions of learning.

The importance of individualized instruction to address this diversity is undisputed in educational theory and research \citep{bernackiSystematicReviewResearch2021}, however, differentiated instruction varies substantially between implementation methods \citep{van_vijfeijken_quantitative_2023}. Translating the theoretical principles into the classroom remains a significant challenge. Teachers face a persistent gap between the pedagogical ideal of differentiated instruction and the practical realities of implementation \citep{robertboschstiftungDeutschesSchulbarometerBefragung2025}. Most available teaching resources are designed for an ``average'' learner and rarely accommodate the diverse cognitive, motivational, and emotional profiles found in today's classrooms, leaving teachers to struggle to create personalized materials within existing time and resource constraints \citep{mollickUsingAIImplement2023}, leading to overall lower quality. 
\citet{pietschEvaluationUnterrichtsstandards2010} proposed a model that incorporates the learning environment, efficiency, student motivation, and differentiation as interconnected elements of effective instruction. This framework acknowledges that successful differentiation must address both cognitive and motivational dimensions---a principle supported by decades of research on intrinsic motivation and self-determination theory \citep{Deci2000-pj}. Therefore, it is insufficient to create learning material that simply recognizes performance outcomes. To effectively support diverse learners, education must also take into account motivational constructs, including intrinsic motivation and self-concept, since these elements are critical in fostering sustained engagement and learning behaviors.

AI-based educational systems have emerged as a promising solution to bridge the theory-practice gap through automated content personalization. 
AI-driven educational platforms are increasingly used worldwide to adapt learning experiences to individual student needs. In the United States, \textit{Khan Academy’s Khanmigo} \citep{khanacademyKhanAcademy2025} employs large language models to provide personalized guidance and adaptive exercises. In China, \textit{Squirrel AI} \citep{squirrelaiSquirrelAI2025} uses AI-powered diagnostic systems to design highly individualized learning trajectories. 
In Singapore, the EdTech Masterplan 2030, as part of Singapore's National Artificial Intelligence (AI) Strategy, supports AI-based learning tools such as adaptive learning systems for the education of students.
In Germany, \textit{alea.schule} \citep{dipfAleaschulePlattformFur2025} offers formative assessment tools that adapt to student performance, while publishers such as \textit{Cornelsen} \citep{cornelsenverlaggmbhKIToolboxCornelsenai2025} integrate AI features for content adaptation. Adoption rates reflect this global momentum: for example, in Germany, 58\% of teachers report using LLMs to prepare teaching materials, with 65\% citing significant benefits for personalizing learning experiences \citep{robertboschstiftungDeutschesSchulbarometerBefragung2025}.

While research has shown that single LLM systems can generate performance-adapted content, they lack the stability and pedagogical depth needed for comprehensive learner modeling that encompasses both psychological and knowledge facets. Moreover, their sensitivity to prompt variations raises concerns about generalizability and highlights the burden placed on non-expert users, such as educators, who must invest substantial effort to ensure high-quality outcomes without extensive prompt engineering \citep{mannekoteCanLLMsReliably2025}. 

More broadly, most AI-based educational systems, LLM-driven or otherwise, remain narrowly performance-focused, adapting content based on quiz results or task completion metrics (e.g. \citealt{xu_ai-driven_2025,lvGenALGenerativeAgent2025}). This limited scope reinforces the theory-practice gap by neglecting the complex diversity that teachers encounter in classrooms. Despite evidence from learning research that effective personalization must integrate motivational, emotional, and self-concept factors that critically influence learning outcomes, many approaches continue to focus narrowly on cognitive aspects alone \citep{bernackiSystematicReviewResearch2021}. 

Multi-agent LLM systems offer a solution to these limitations by moving beyond single-prompt approaches toward coordinated systems of specialized agents. Unlike single-prompt approaches, such architectures enable the creation of specialized, reusable learner profiles that maintain consistency across interactions while representing both psychological and cognitive dimensions \citep{razafinirinaPedagogicalAlignmentLarge2024,wang_llm-powered_2025}. Recent studies further demonstrate that LLMs can embody stable personality traits and motivational stances with high internal consistency \citep{huang_designing_2026,wang_evaluating_2025}, and exhibit systematic patterns across motivation, affect, and behavioral dimensions \citep{li_quantifying_2024}.

\subsection{Related work} 
\label{sec:relwork}
\paragraph{Principles of Effective Personalization}
First, effective personalization requires modeling both cognitive (prior knowledge, skills) and motivational aspects (motivation, emotional disposition, self-efficacy) \citep{bernackiSystematicReviewResearch2021}. According to Self-Determination Theory, supporting autonomy, competence, and relatedness is essential for learning success \citep{Deci2000-pj}. 
It has been understood that the implementation of personalized learning can improve the learning outcomes as well as increase the motivation \cite{bernackiSystematicReviewResearch2021,farianiSystematicLiteratureReview2023}. 
In mathematics, emotional factors like anxiety and self-concept are critical for engagement and persistence, making motivational adaptation as vital as cognitive scaffolding \citep{alamriUsingPersonalizedLearning2020,darling-hammondImplicationsEducationalPractice2020}.

Second, adaptive instructional design must go beyond content selection to include instructional strategies, scaffolding, and feedback, all of which are vital for learner engagement and self-regulation \citep{xuEffectsTeacherRole2020}. Scaffolding—such as examples and stepwise guidance—supports learners’ progression toward independence. Context-sensitive personalization, which incorporates student interests, real-world relevance, and situational constraints, is essential for maintaining authenticity \citep{changSystematicReviewTrends2020a}. This is especially important in mathematics, where self-concept, intrinsic motivation, and prior knowledge strongly influence outcomes \citep{alamriUsingPersonalizedLearning2020,darling-hammondImplicationsEducationalPractice2020}. Providing agency, competence-matched activities, and supportive social contexts increases engagement and meaning, making it crucial that individualization addresses emotional and motivational as well as cognitive needs.
Third, pedagogical intentionality ensures personalization is aligned with learning goals and grounded in educational theory \citep{bernackiSystematicReviewResearch2021}. Bloom’s revised taxonomy offers a structured framework for guiding learners from basic recall to higher-order thinking \citep{andersonTaxonomyLearningTeaching2001}, enabling educators and adaptive systems to scaffold experiences that build on current understanding and foster progression to complex reasoning.

\paragraph{LLMs for Simulating Diverse Personas}
A key foundation of our approach is the demonstrated ability of LLMs to simulate complex individual characteristics with high fidelity. Recent research shows LLMs can reliably embody stable personality traits and motivational dispositions, achieving internal consistency and convergent validity on par with or exceeding human responses \citep{aher_using_2023, jiang_personallm_2024}. LLMs also reflect human-like cognitive biases in reasoning, adding psychological realism to their behavioral simulations \citep{lampinenLanguageModelsHumans2024,ozekiExploringReasoningBiases2024}. This enables the creation of learner agents with coherent motivational stances and learning behaviors over extended interactions.
Crucially, LLMs overcome the scalability limitations of traditional simulated learners, which required extensive manual development \citep{blessingEvaluatingAuthoringTool2008}, by enabling rapid, automated generation of diverse, reusable learner profiles. Building on these capabilities, recent work has introduced LLM-based ``digital twins'' for student behavior and multi-agent systems for coordinated pedagogical scenarios \citep{xuClassroomSimulacraBuilding2025}. However, most of these applications typically remain student-facing rather than directly supporting teachers \citep{xuEduAgentGenerativeStudent2024}.

\paragraph{The Need for Teacher-Facing Support Systems}
Despite significant progress in student-facing applications \citep{wang_llm-powered_2025, parkEmpoweringPersonalizedLearning2024}, most current LLM systems do not provide direct support for teachers. Although their support is essential to ensure curriculum alignment and pedagogical coherence in physical classroom environments and research shows concrete touch points for using LLMs throughout the teaching process \citep{schummel_specifying_2025}. Existing teacher-facing systems primarily offer static content repositories or basic customization features, requiring significant manual effort to tailor materials for individual students \citep{xuEduAgentGenerativeStudent2024}.

Recent work has begun exploring LLM-based teacher support, including the generation of a lesson plan and the development of instructional strategies \citep{huExploringPotentialLLM2025,zhang_simulating_2025}. These systems can assist in producing high-quality teaching plans and offer substantial potential to reduce planning workload by simulating students with diverse needs, enabling proactive identification and resolution of potential task difficulties \citep{xuEduAgentGenerativeStudent2024}. However, these approaches lack the systematic integration of learner simulation, pedagogical frameworks, and multi-dimensional individualization that would enable comprehensive support for classroom personalization. Multi-agent architectures offer particular promise for teacher-facing systems by enabling specialized roles, reusable learner profiles, and coordinated pedagogical reasoning within a ``human-in-the-loop'' model where teachers maintain control over AI tools \citep{chu_llm_2025}, yet remain underexplored for comprehensive classroom support.

\subsection{Contribution}
Building on these advances, we introduce a multi-agent system designed to overcome the narrow performance focus of existing personalization approaches. Our framework employs learner agents that model psychological (e.g., motivation, self-concept) and knowledge dimensions, teacher agents that incorporate curriculum-specific and pedagogical criteria, and evaluator agents that provide automated quality assurance. Thereby, our framework architecture delivers stable and reproducible outputs via persistent agent profiles, reduces prompting overhead through reusable student-specific agents,  integrates school- and subject-specific pedagogical requirements, and supports semi-automated generation of teaching materials with built-in quality control.

Our paper advances the emerging research agenda on generative tools for inclusive, adaptive teaching by making four key contributions: (1) We design and implement a teacher-facing, LLM-based multi-agent system for classroom personalization (Figure~\ref{fig:banner}), (2) we integrate this system with authentic curriculum content using authentic tasks from the 8th-grade mathematics curriculum in German schools to ensure relevance and practical applicability, (3) we evaluate the system through teacher feedback and scenario-based implementation in realistic contexts; and (4) we demonstrate the feasibility and pedagogical potential of multi-agent LLM systems for holistic personalization that addresses both cognitive and motivational learner dimensions in physical classrooms. 

\section{Problem Statement} 
\label{sec:probl}

Despite the growing integration of AI in education, current personalization systems face two major limitations: (1) They are predominantly restricted to performance-based adaptations within digital environments, overlooking the complex, multidimensional needs of learners in physical classrooms. Such needs are shaped by motivation, self-concept, cultural background, language proficiency, and neurodiversity; (2) Existing systems are largely student-oriented, providing limited direct support to teachers tasked with differentiating instruction for increasingly heterogeneous classrooms. Few AI tools engage with the didactical challenges of tailoring content across diverse learner profiles.

Addressing these challenges requires modular, teacher-oriented AI systems capable of simulating realistic learner characteristics and generating instructional materials aligned with established educational frameworks. In this paper, we present \facet, a modular agent-based framework designed to help teachers personalize learning materials by simulating the engagement of learners. We investigate the following research question: 

\begin{center}
    \textit{Can a multi-agent framework generate teaching materials that are effectively adapted to individual learner needs? }
\end{center}

In this paper, we present results on the quality and suitability of the generated materials using two complementary methods: automated analysis using LLMs and preliminary, exploratory feedback from K-12 in-service teachers.

\section{The \facet Framework}
\label{sec:system}
\begin{figure*}[htbp]
    \centering
    \includegraphics[width=1\linewidth]{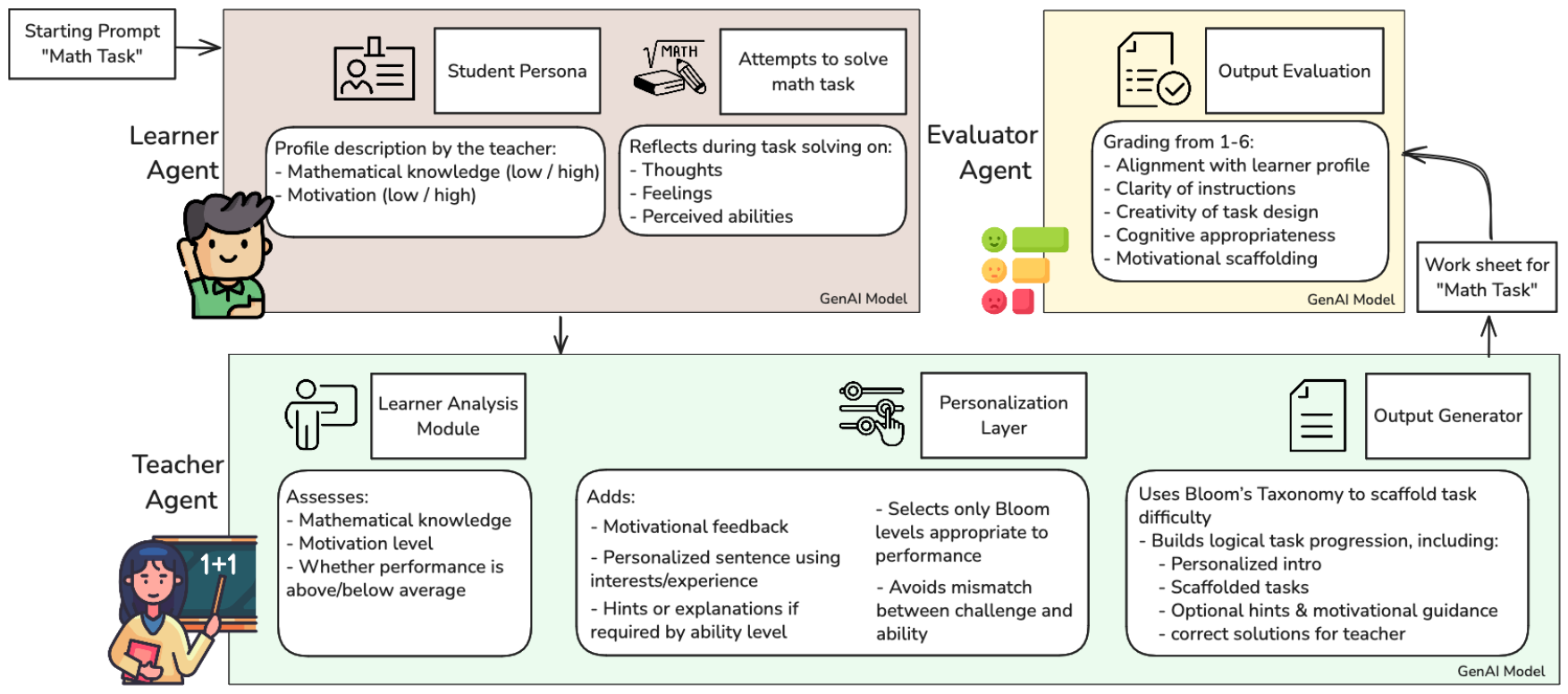}
    \caption{Modular agent architecture of the \facet Framework}
    \caption*{\textit{Note: Learner agents represent distinct cognitive and motivational profiles. The teacher agent assesses these profiles and generates personalized worksheets based on didactical principles. The evaluator agent reviews the resulting instructional materials for quality and alignment with the learner profiles.}}
    \label{fig:agent_architecture}

\end{figure*}

\textit{\facet}, a \textbf{F}ramework for \textbf{A}gent-based \textbf{C}lassroom \textbf{E}nhancement for \textbf{T}eacher, is designed to simulate and analyze learner behavior, with the objective of personalizing output, such as worksheets. An evaluation, integrated into the system architecture, ensures the quality of the personalization. The framework has four key characteristics: (1) It is \textit{agent-based}, which increases the robustness of the learner personas compared to single-LLM setups and reduces prompting effort by allowing teachers to define learner personas once and instantiate diverse variants of learner agents with minimal additional input; it includes (2) \textit{learner agents} characterized by diverse dimensions, encompassing performance and motivational factors, and (3) a \textit{teacher agent}, that provides output considering performance, instructional aspects, such as real-world relevance, motivational support, scaffolding, and hints, that explicitly address learner motivational aspects, next to performance; (4) an integrated \textit{evaluator agent}, that provides feedback on the suitability of the output with respect to the diverse learners' needs. 
Figure~\ref{fig:agent_architecture} illustrates the proposed framework, detailing the workflow used to generate individualized worksheets, which is exemplified using the knowledge and motivation of students as characteristics of learners.

\paragraph{Multi-Agent System} The modular design of the \facet Framework reflects a teacher-oriented model of personalization. By embedding personalization decisions in separate, transparent agent modules, our system supports experimentation, refinement, and ultimately more effective classroom practices.

From a technical perspective, the system instantiates three LLM-backed agents with explicit roles and typed interfaces. After piloting several models, we use GPT-4.1 for the reasoning-heavier learner and teacher agents, and GPT-4o for the evaluator and lightweight formatting, which yielded stable performance with acceptable latency. Each agent is driven by a reusable prompt template with slot-filling for grade, topic, and profile parameters (listed in the Appendix). Our system supports more than 100 LLMs via LiteLLM \citep{berriaiLiteLLM2025}, including models from OpenAI, xAI, and Anthropic. Inter-agent communication is strictly structured: the learner agent emits a transcript capturing attempted solution steps, misconceptions, and affective cues; the teacher agent consumes this transcript and produces a one-page worksheet object with tasks annotated by Bloom level, scaffolds, hints, and motivational comments; the evaluator returns rubric-aligned scores and concise diagnostics. To improve reproducibility, we encode hard constraints in prompts (e.g., task count, length budget). The pipeline executes sequentially (learner \(\rightarrow\) teacher \(\rightarrow\) evaluator); we tested several iterative refinement schemes, but additional iterations did not improve the results; a single iteration already yielded sufficiently accurate answers. The used modularization reduces prompting overhead for teachers, preserves the interpretability of intermediate decisions, and supports the stable behaviors observed in our evaluations; prompt templates for the teacher and evaluator agents are listed in the Appendix. The teacher accesses the system via a custom web interface. 

\paragraph{Simulating Diverse Learner Agents}
The \facet Framework provides a scalable and controllable multi-agent system for simulating diverse learning personas by incorporating cognitive, affective, and behavioral variables. A learner persona can be instantiated through prompting to reflect diverse characteristics. The learner agent is exposed to a teacher-defined task prompt and attempts to solve the task while generating self-reflective output that captures its reasoning process, emotional state, and perceived self-efficacy. Leveraging an LLM's capacity to simulate individual characteristics enables the adaptation of instructional content to individual variation and supports teacher-facing personalization informed by realistic learner behavior, as indicated by \citet{li_quantifying_2024}.

\paragraph{Including Didactical Directives via Teacher Agent}
The teacher agent can be prompted with high-level didactical directives that define the instructional orientation of the generated worksheet. These prompts may specify preferences regarding instructional style, motivational theory, or task selection strategies, allowing the system to align with diverse didactic goals and classroom contexts. The teacher agent comprises three interrelated aspects designed to adapt instruction to individual learner profiles; this can be understood as a set of meta-prompting strategies in which content generated by the learner agent informs the teacher’s subsequent actions, as described in the following three aspects:

\textit{Analysis module} This module performs a structured evaluation of the learner model, integrating indicators of mathematical proficiency and motivational disposition. Identifies deviations from normative benchmarks, providing a diagnostic basis for subsequent adaptation.

\textit{Personalization layer} This layer consists of two aspects: Based on the analysis, this layer, firstly, supports the generation of adaptive instructional responses. These include motivational feedback, content customization based on interests and scaffolding prompts aligned with the learner's prior knowledge and self-reported dispositions, and, secondly, personalized task selection by mapping learner profiles to stratified tasks, e.g., along Bloom's taxonomy. Logical alignment ensures cognitive alignment by selecting tasks that challenge the learner without exceeding their current developmental capacity.

\textit{Output generator} This component compiles the system's personalized output as a coherent worksheet formatted in Markdown and thereby integrates tasks, feedback, and instructional support into a structured artifact designed to be delivered through teacher-facing web interfaces.

\paragraph{Output Feedback Using the Evaluator Agent} 
The evaluation component is designed to provide structured feedback on the effectiveness of instructional tasks in relation to diverse learner profiles. Firstly, this plays a central role in improving the alignment between instructional design and individual learning needs, thus forming iterative improvements to the system. Secondly, this feedback is valuable for teachers who have typically not been trained in prompt engineering, as it provides feedback on inconsistencies between the learners' prompts and the intended output. At its core, the evaluator agent provides feedback along four key dimensions: (1) \textit{Didactical structure} assesses the didactical soundness and coherence of the task design, (2) \textit{Clarity of the task} evaluates the extent to which task instructions and expectations are comprehensible to learners, (3) \textit{Creativity} reflects the originality and engagement potential of the task, particularly in relation to maintaining learner interest, and (4) \textit{Suitability for learners' needs} judges the appropriateness of the task in terms of matching the learner's cognitive and motivational profile. 

\section{Evaluation Method}
We evaluated the feasibility of the proposed \facet framework and examined whether a multi-agent architecture can generate teaching materials effectively adapted to individual learners’ needs. The evaluation aims to validate the core system’s functionalities and output through an LLM-based evaluation (internal validity) and to assess real-world applicability and practical relevance via preliminary exploratory feedback from K-12 in-service teachers (external validity).

\paragraph{Learning Scenario} 
We situated our evaluation within the context of nationally standardized curriculum frameworks. Specifically, instructional tasks were derived from curriculum standards for grade 8 mathematics in German secondary education. We focus on Germany as a concrete example—a country with education metrics close to the OECD standard according to the 2022 PISA assessment \citep{oecdPISA2022Results2023}, growing heterogeneity among students similar to many Western countries, and clear standards for each school year, which makes it representative of many school formats. To start the conversation, the learner agents engage in a starting prompt that includes mathematical tasks for grade 8, such as: \textit{'Using the digits 1, 2, 3, 4, and 5, form four-digit numbers where no digit is repeated. The first digit of the number must be greater than 3, and the second digit must be even. How many different four-digit numbers can be created with these restrictions?'}

\paragraph{Learner and Teacher Agent Representation}
For our experiments, we conceptualized learners through two core dimensions: intrinsic motivation and mathematical proficiency, reflecting the learner's subject-specific knowledge. These constructs are defined theoretically as follows:

\textit{Intrinsic motivation} refers to engaging in an activity for its inherent satisfaction and interest, rather than for external rewards or pressures (see Self-Determination Theory by \citet{Deci2000-pj}). 
Intrinsic motivation was prompted as ``high'' or ``low'' based on the standardized PISA scale, in which students responded to items reflecting four dimensions: interest/enjoyment, perceived competence, perceived choice, and pressure/tension \citep{bosLernmotivationMathematikSchuler2012}. In the literature (see e.g., \citet{Deci2000-pj}), the magnitude of motivation is described as a continuum. In our initial model, however, we distinguish only between ``high'' and ``low'' motivation. The framework can nevertheless accommodate finer-grained distinctions beyond this binary categorization. \textit{Mathematical proficiency}, as the ability to understand and solve mathematical problems effectively. Mathematical knowledge was initially included in the learner prompt as ``high'' or ``low''. The students then engaged in a given starter prompt (mathematical task) aligned with their current level of knowledge. 

Each agent functions according to a customized instructional prompt and is guided to engage in mathematical tasks in a manner that aligns with its assumed cognitive abilities and motivational profile. 
We defined a set of four learner profiles to test the framework's adaptability across cognitive and motivational characteristics. We are aware that other individual factors like gender might potentially bias the LLMs response (for a review, see \citealt{nemani_gender_2024}), which could potentially impact the way LLMs individualize learning material. As such an analysis is beyond the scope of this paper, the learners’ gender was held constant, with only mathematical proficiency and motivation varied (see the learner prompt in the Appendix).

The teacher agent operated under high-level didactical directives defining the instructional focus, motivational framework, and task selection, in alignment with Bloom’s taxonomy. Its modular structure includes an analysis module to assess learner profiles, a personalization layer to generate adaptive instruction and select tasks, and an output generator that compiles a coherent one-page worksheet with personalized content, scaffolding, and motivational support (see Appendix).

\paragraph{Agent-Based Evaluation}
To assess the quality of personalized output, we implemented an automated evaluation pipeline using the third agent as an evaluator agent. This agent receives the personalized worksheet as input and produces structured feedback on the four previously defined dimensions: didactical structure, clarity of the task, creativity, and suitability for learners' needs. Each dimension is operationalized through a 1 (low) to 6 (high) ranking, with clearly defined descriptors to score the quality and suitability of the worksheets. A summary of the criteria for each dimension is listed in the Appendix.  

\paragraph{Expert-Based Evaluation}
In addition, we gathered preliminary, exploratory feedback from mathematics K-12 in-service teachers. We conducted a comparative analysis between two types of worksheets: (1) the personalized worksheet, automatically generated by \facet Framework based on learner-specific profiles (mathematical knowledge, motivation), to (2) a baseline worksheet, representing a standard, non-adaptive instructional design typical of conventional classroom practice. The evaluation followed a structured three-step protocol with the teachers: Participants were asked to review all four personalized worksheets, verbalizing their feedback. Each worksheet was rated on the four dimensions, with teachers providing a 1 (low) to 6 (high) score and open-ended feedback on the following questions:
What aspects of the output do you find valuable? What is missing to improve differentiation, motivation, and clarity? 

\section{Results and Discussion} 
\label{sec:results}
We present an example of a generated personalized worksheet and its evaluation, based on ten agent-based iterations per learner profile and expert feedback from five K–12 in-service teachers, including ratings and qualitative insights for system refinement.

\begin{table*}[t]
\small
\centering
\renewcommand{\arraystretch}{1.3}
\captionsetup{labelfont={it}, textfont={it}}
\caption{Results of Worksheet Evaluation: Presented in terms of internal validity (agent-based evaluation) and external validity (teacher expert evaluation) – inverted 6-point scale (1 = insufficient, 6 = very good)}
\label{tab:agent_expert_ranking_inverted}
\resizebox{\textwidth}{!}{%
\begin{tabular}{p{2.3cm}cc cc cc cc}
    \toprule
\textbf{Dimension} & \multicolumn{2}{c}{\(K_\mathrm{L}\; M_\mathrm{H}\)} & \multicolumn{2}{c}{\(K_\mathrm{L}\; M_\mathrm{L}\)} & \multicolumn{2}{c}{\(K_\mathrm{H}\; M_\mathrm{L}\)}  & \multicolumn{2}{c}{\(K_\mathrm{H}\; M_\mathrm{H}\)}  \\
    \cmidrule(lr){2-3} \cmidrule(lr){4-5} \cmidrule(lr){6-7} \cmidrule(lr){8-9}
    & Agent & Teacher Expert & Agent & Teacher Expert & Agent & Teacher Expert & Agent & Teacher Expert \\
    \midrule
\textbf{Didactical structure} & 5.1 (0.3) & 4.7 (4-6) & 5 (0) & 5 (4-6) & 5.1 (0.3) & 5.1 (3-6) & 5.4 (0.5) & 5 (4-6) \\
\textbf{Clarity of the task} & 5.9 (0.3) & 4.3 (3-6) & 6 (0) & 4.8 (3-6) & 5.9 (0.3) & 4.9 (3-6) & 6 (0) & 4.7 (3-6) \\
\textbf{Creativity} & 4.9 (0.3) & 4.7 (3-6) & 4.9 (0.3) & 4.4 (4-5) & 5.1 (0.3) & 4.4 (3-6) & 4.8 (0.4) & 4.6 (3-6) \\
\textbf{Suitability of the task for learners} & 6 (0) & 4.8 (4-6) & 6 (0) & 4.9 (4-6) & 6 (0) & 5.1 (5-6) & 5.9 (0.3) & 5.4 (5-6) \\
\bottomrule
\end{tabular}
}
\vspace{0.5em}
\begin{flushleft}
{\footnotesize 
\textit{Note}: Internal validity was evaluated by performing 10 iterations per worksheet, and external validity was evaluated with five teacher experts. Rankings are presented as mean values \textit{(M)} with standard deviation \textit{(SD)} for agent-based evaluation and \textit{(range)} for teacher-based evaluation. Evaluation based on an inverted $6$-point scale ($1$ = insufficient, $6$ = very good). The four example learner groups are:  
\(K_\mathrm{L}\;M_\mathrm{H}\) = Low knowledge, high motivation;  
\(K_\mathrm{L}\;M_\mathrm{L}\) = Low knowledge, low motivation;  
\(K_\mathrm{H}\;M_\mathrm{L}\) = High knowledge, low motivation;  
\(K_\mathrm{H}\;M_\mathrm{H}\) = High knowledge, high motivation.}
\end{flushleft}
\end{table*}

\paragraph{Agent-Based Evaluation} The results demonstrate that the proposed multi-agent framework can generate teaching materials tailored to individual learner profiles. As shown in Table~\ref{tab:agent_expert_ranking_inverted}, the agent-based evaluation returned consistently high ratings across all four profiles and ten iterations, underscoring the \facet Framework’s stable performance in generating reliable worksheet outputs (a table presenting all metrics for ten iterations is listed in the Appendix). 
The strongest scores were observed for suitability of the task ($\mathit{M} = 4.78$, $\mathit{SD} = 0.06$) and clarity of the task ($\mathit{M} = 4.76$, $\mathit{SD} = 0.12$) for all four learner personas, indicating very high internal validity. Didactical structure ($\mathit{M} = 4.12$, $\mathit{SD} = 0.22$) and creativity ($\mathit{M} = 3.94$, $\mathit{SD} = 0.26$) were also rated favorably, with slightly lower but still strong scores. These findings suggest that the generated worksheets are both pedagogically sound and well-aligned with learner needs.

Each final output consists of a personalized worksheet with a clear task sequence aligned to Bloom’s taxonomy and the learner’s profile \citep{andersonTaxonomyLearningTeaching2001}. Task difficulty is dynamically selected to match the learner's current performance level, avoiding both under- and over-challenge. To achieve this, tasks are drawn from three cognitive categories: reproduction (recall and procedure), making connections (application in novel but related contexts), and generalizing and reflecting (abstraction and conceptual reasoning). The worksheet begins with a personalized introduction connecting to the learner’s interests or prior experiences, followed by tasks in logical progression. Each task is accompanied by guiding comments responsive to the learner’s motivational state, and hints or short explanations are provided where knowledge gaps are detected. (Figure~\ref{fig:exampleoutput} illustrates an example worksheet for a student with limited knowledge and high motivation to learn.

\paragraph{Teacher Expert-Based Evaluation}
To evaluate the real-world applicability and practical relevance of the generated output, we collected exploratory feedback from five K–12 in-service teachers ($4$ male, $1$ female; $\mathit{M} = 38$ years, $\mathit{SD} = 6.6$), averaging $6.4$ years of teaching experience ($\mathit{SD} = 2.7$). Overall, the results indicate a high perceived quality of the personalized worksheets. The dimension didactical structure ($\mathit{M} = 4.95$, $\mathit{range} = 3-6$) and suitability of the task for learners were rated positively in all profiles ($\mathit{M}  = 5.05$, $\mathit{range} = 4-6$), indicating that the personalization logic effectively aligned task demands with each learner’s cognitive and motivational characteristics. In contrast, ratings for clarity of the task ($\mathit{M} = 4.68$, $\mathit{range} = 3-6$) and creativity ($\mathit{M} = 4.53$, $\mathit{range} = 3-6$) remained favorable overall. 

These patterns were reinforced by qualitative feedback obtained via think-aloud protocols. For learners with limited knowledge and low motivation, teachers suggested several refinements: clearer instructional scaffolding, starting worksheets with explicit mathematical operators, integrating worked examples, and adding guided solution strategies such as prefilled rows or structured hints. For students with low motivation, experts recommended deepening personalization by embedding narrative elements and everyday-life relevance to foster engagement and problem-solving motivation. Teachers also appreciated the reduced number of tasks for students with low motivation, noting that it prevented cognitive overload. Overall, the evaluations suggest that the K-12 in-service teachers value the opportunity to personalize worksheets for heterogeneous classrooms. Two representative quotations below illustrate individual perspectives: \newline

\textit{``very cool tool as I can adjust tasks in the difficulty---which means the whole class `stays' within one topic, while no one is bored, while prior the tool I had do move the fast and motivated students to the new topics, which created a speed imbalance in the class -- now, I can satisfy my really good students with just much more complex tasks''} (ID 2) \newline

\textit{``I would love to have a `Test-preparation kid' for exams, so that I can try solving the exam once in advance and check the clarity of the tasks and the likelihood of students being able to solve them, as well as create two versions that can be created with identical difficulty.''} (ID 2) \newline

The preliminary results present several limitations that inform future research. Firstly, the teacher evaluation was conducted outside of authentic classroom contexts, making it difficult to fully assess the suitability of generated tasks for specific learners. In the next study, we plan to test our \facet framework in large-scale, in-classroom evaluations using outputs tailored to real learner profiles. Secondly, in the current prototype, learner agents were prompted with sparse input to test potential and robustness to adapt learning material. While this minimal prompting yielded a notable degree of personalization, it does not capture the full diversity of real classrooms. Our future vision is to enrich learner profiles with additional dimensions such as language background, attentional variability, and neurodiversity (e.g., ADHD, dyslexia). This scalable approach would enable the generation of realistic learner personas that support authentic, individualized instructional design. Future system versions will enable teachers to represent diverse classes by generating tailored learner personas for each student, informed by diverse characteristics. Thirdly, based on teacher-expert feedback, we intend to extend teacher agents to incorporate operators and motivational strategies grounded in educational theory, and embed full curriculum specifications into prompts. This would align generated materials with established standards, improve pedagogical fit, and ensure domain-appropriate task generation.

\section{Conclusion}
This study presents preliminary results demonstrating the feasibility of the \textit{\facet framework}, a multi-agent AI architecture, to generate individualized teaching materials that align with diverse learner needs. By modeling learners along cognitive and motivational dimensions and aligning outputs with established educational principles, the framework produced stable, high-quality worksheets, as confirmed through agent-based evaluation and K-12 in-service teacher feedback. While the current implementation relies on intentionally minimal prompting, the results highlight the adaptability of LLM-based multi-agent systems for classroom personalization. Future work should enrich learner and teacher agent profiles, integrate curricular constraints and context-specific didactic theories, and conduct in-situ classroom evaluations to validate effectiveness under authentic conditions. Ultimately, this approach offers a scalable, learner-sensitive pathway for AI-assisted teacher support, advancing the research agenda on generative tools for inclusive and adaptive education.

\section{Acknowledgments}
We would like to thank the Zuse Institute Berlin (\url{https://www.zib.de}) for hosting various LLM models for testing. Research reported in this paper was partially supported by the German Federal Ministry of Research, Technology and Space, grant numbers 16DII133 and 16DII137 (Weizenbaum-Institute).

\bibliographystyle{aaai2026} 
\bibliography{references}

\newpage
\appendix

\section{Appendix A}

\label{sec:App_A}
\begin{figure*}
    \centering
    \includegraphics[width=0.85\linewidth]{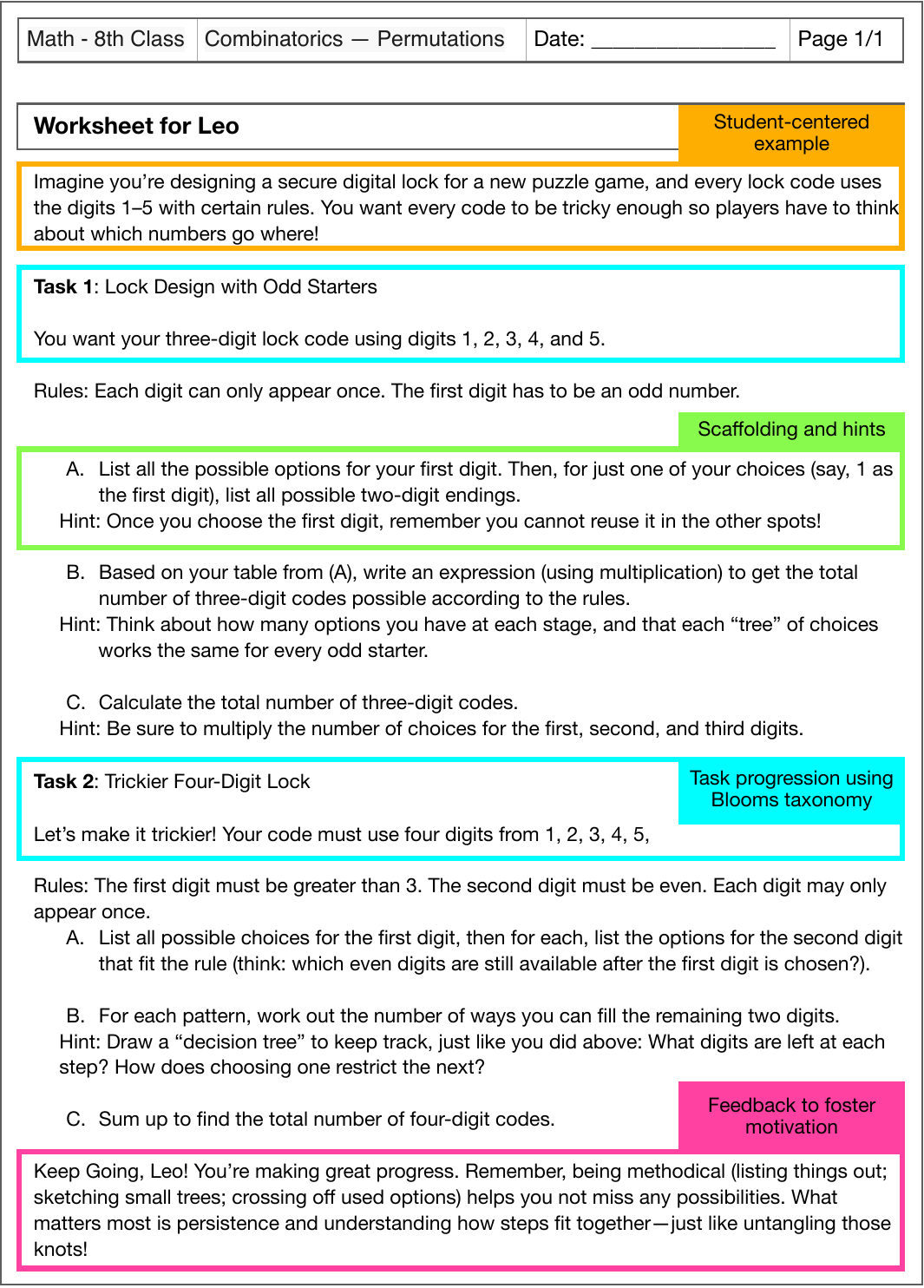}
    \caption{Components of a personalized worksheet (colored boxes are not printed on the original worksheet; they are added here to highlight the conceptual components)}
    \label{fig:exampleoutput}
\end{figure*}

Example of a generated personalized worksheet using the \facet framework, tailored to a learner who demonstrates high motivation but possesses limited subject knowledge. The worksheet begins with an interest-based introductory sentence to foster engagement, followed by a logically sequenced set of tasks aligned with the learner’s performance level. Each task includes guiding comments adapted to the learner’s profile and, when needed, hints or explanations to support comprehension. Figure~\ref{fig:exampleoutput} illustrates this personalized output.

\section{Appendix B}
\label{sec:App_B}

Appendix B presents the prompts for the learner agent, teacher agent and the evaluator agent used in the evaluation. 

\subsection{Learner agent prompt}
\begin{lstlisting}[style=feedbackprompt]
You are a 15-year-old male student with above-average mathematical performance, and strong motivation to learn
You work on tasks that align with your current level of knowledge, self-perception, and motivation. You reflect on your thoughts, feelings, and mathematical abilities throughout the process
    \end{lstlisting}

\subsection{Teacher agent prompt}
\begin{lstlisting}[style=feedbackprompt]
You provide a clear picture of the student's learning situation, taking into account the mathematical knowledge and motivation, and identify and explain whether the student is performing below or above average.
You present the tasks on a one-page worksheet. The worksheet starts with a personalized sentence that draws on the student's interests and experiences, making it an engaging experience for them. From there, you design new tasks based on Bloom's Taxonomy (Blütenaufgaben) and you select the task difficulty in alignment with the student's current performance level, using only levels that appropriately match the student's abilities. Tasks that are too easy or too difficult are deliberately avoided.
easy: Reproduction---recall and apply learned procedures.
medium: Making Connections---apply knowledge in new but related contexts.
advanced: Generalizing and Reflecting---abstract reasoning and reflection on concepts.
The tasks should build logically upon one another, following a clear progression. Each task should be aligned with Bloom's Taxonomy ('Blütenaufgaben') and tailored to the student's current performance level, ensuring that tasks are appropriately challenging without being too easy or too difficult.
If the student's knowledge level requires it, add hints and brief explanations to support understanding and reflect the student's current level of knowledge.
If the abilities or motivation level of the student requires it, add personalized comments that guide the student through the worksheet and acknowledge their level of motivation.
You present the tasks on a one-page worksheet. The worksheet starts with a personalized sentence that draws on the student's interests and experiences, making it an engaging experience for them. From there, you design new tasks based on Bloom's Taxonomy ('Blütenaufgaben') and you select the task difficulty in alignment with the student's current performance level, using only levels that appropriately match the student's abilities. Tasks that are too easy or too difficult are deliberately avoided.
easy: Reproduction---recall and apply learned procedures.
medium: Making Connections---apply knowledge in new but related contexts.
advanced: Generalizing and Reflecting---abstract reasoning and reflection on concepts.
\end{lstlisting}

\subsection{Evaluator agent prompt}
\begin{lstlisting}[style=feedbackprompt]
Evaluate Mr. Taylor's worksheet using the following four criteria. 

Criteria 1: Didactical structure = How logically and pedagogically the tasks are organized and sequenced
1 = Highly original or open-ended task allowing for multiple perspectives or solutions; stimulates creative thinking. 
2 = Some creative elements or opportunities for individual approaches. 
3 = Standard task, no special creative value. 
4 = Task forces pseudo-creative framing without real value or makes task harder through unnecessary open-endedness. 
5 = ``Creative'' elements are misleading or inappropriate for the learning goal. 
6 = No creative aspects or task is missing. 

Criteria 2: Clarity of the tasks = How clearly the instructions and questions are formulated
1 = Task is clearly formulated, logically sequenced, easy to understand for the target group. 
2 = Mostly clear with only minor ambiguities; task goal is still understandable. 
3 = No special consideration for clarity; some unclear formulations. 
4 = Overcomplicated or verbose formulation that makes the task harder than necessary. 
5 = Formulation is misleading or includes wrong cues or terminology.
6 = Task is missing or incomprehensible. 

Criteria 3: Creativity of the tasks = The originality and engagement level of the task design.
1 = Highly original or open-ended task allowing for multiple perspectives or solutions; stimulates creative thinking. 
2 = Some creative elements or opportunities for individual approaches. 
3 = Standard task, no special creative value. 
4 = Task forces pseudo-creative framing without real value or makes task harder through unnecessary open-endedness. 
5 = ``Creative'' elements are misleading or inappropriate for the learning goal. 
6 = No creative aspects or task is missing. 

Criteria 4: Suitability of the tasks for learner: Appropriateness of the task in terms of matching the learner's cognitive and motivational profile.
1 = Perfectly adapted to the learners' prior knowledge, interests, and level of difficulty; starts from the concrete/known. 
2 = Well adapted, minor mismatches with regard to content or level. 
3 = No effort made to adapt to learner level (e.g., too generic or average). 
4 = Adaptation attempt made, but misjudges learner needs or overcomplicates. 
5 = Task is developmentally inappropriate or contains false assumptions about learner ability. 
6 = Task not completed or fully unsuitable for intended learners. 

Use a scale from 1 to 6 where 3 represents the quality of an average worksheet. This is an average worksheet:

\end{lstlisting}

\section{Appendix C}
\label{sec:App_C}

Appendix C presents the evaluation metrics used in the agent-based and teacher-based evaluations. 

\begin{table}
\small
\renewcommand{\arraystretch}{1}
\footnotesize
\captionsetup{labelfont={it}, textfont={it}}
\caption{Didactical Structure}
\label{tab:agentfeedback_didacticalstructure}
\begin{tabular}{
    >{\raggedright\arraybackslash}p{0.5cm} 
    >{\raggedright\arraybackslash}p{7cm}}
\toprule
\textbf{Score} & \textbf{Description} \\
\midrule
1 & Clear didactical progression, from simple to complex, from concrete to abstract, and goals are transparent. \\
2 & Mostly clear structure and goal transparency, useful guidance for different levels. \\
3 & No visible adaptation to student perspective. \\
4 & Structure is hard to follow or overly complex without didactical necessity. \\
5 & Confusing or misleading structure; goals unclear; no consolidation or even contradictory instructions. \\
6 & No worksheet or completely unusable structure. \\
\bottomrule
\end{tabular}
\begin{flushleft}
{\footnotesize 
\textit{Note}:Didactical structure assesses the didactical soundness
and coherence of the task design}
\end{flushleft}
\end{table}

\begin{table}
\small
\renewcommand{\arraystretch}{1}
\footnotesize
\captionsetup{labelfont={it}, textfont={it}}
\caption{Clarity of the Task}
\label{tab:agentfeedback_clarity}
\begin{tabular}{
    >{\raggedright\arraybackslash}p{0.5cm} 
    >{\raggedright\arraybackslash}p{7cm}}
\toprule
\textbf{Score} & \textbf{Description} \\
\midrule
1 & Task is clearly formulated, logically sequenced, easy to understand for the target group. \\
2 & Mostly clear with only minor ambiguities; task goal is still understandable. \\
3 & No special consideration for clarity; some unclear formulations. \\
4 & Overcomplicated or verbose formulation that makes the task harder than necessary. \\
5 & Formulation is misleading or includes wrong cues or terminology. \\
6 & Task is missing or incomprehensible. \\
\bottomrule
\end{tabular}
\begin{flushleft}
{\footnotesize 
\textit{Note}: Clarity of the task evaluates the extent to which task in-
structions and expectations are comprehensible to learn-
ers.}
\end{flushleft}
\end{table}

\begin{table}
\small
\renewcommand{\arraystretch}{1}
\footnotesize
\captionsetup{labelfont={it}, textfont={it}}
\caption{Creativity}
\label{tab:agentfeedback_creativity}
\begin{tabular}{
    >{\raggedright\arraybackslash}p{0.5cm} 
    >{\raggedright\arraybackslash}p{7cm}}
\toprule
\textbf{Score} & \textbf{Description} \\
\midrule
1 & Highly original or open-ended task allowing for multiple perspectives or solutions; stimulates creative thinking. \\
2 & Some creative elements or opportunities for individual approaches. \\
3 & Standard task, no special creative value. \\
4 & Task forces pseudo-creative framing without real value or makes task harder through unnecessary open-endedness. \\
5 & `Creative' elements are misleading or inappropriate for the learning goal. \\
6 & No creative aspects or task is missing. \\
\bottomrule
\end{tabular}
\begin{flushleft}
{\footnotesize 
\textit{Note}: Creativity reflects the originality and engagement po-
tential of the task, particularly in relation to maintaining
learner interest.}
\end{flushleft}
\end{table}

\begin{table}
\small
\renewcommand{\arraystretch}{1}
\footnotesize
\captionsetup{labelfont={it}, textfont={it}}
\caption{Suitability of the Task for Learners}
\label{tab:agentfeedback_suitability}
\begin{tabular}{
    >{\raggedright\arraybackslash}p{0.5cm} 
    >{\raggedright\arraybackslash}p{7cm}}
\toprule
\textbf{Score} & \textbf{Description} \\
\midrule
1 & Perfectly adapted to the learners' prior knowledge, interests, and level of difficulty; starts from the concrete/known. \\
2 & Well adapted, minor mismatches with regard to content or level. \\
3 & No effort made to adapt to learner level (e.g., too generic or average). \\
4 & Adaptation attempt made, but misjudges learner needs or overcomplicates. \\
5 & Task is developmentally inappropriate or contains false assumptions about learner ability. \\
6 & Task not completed or fully unsuitable for intended learners. \\
\bottomrule
\end{tabular}
\begin{flushleft}
{\footnotesize 
\textit{Note}: Suitability for learners' needs judges the appropriateness of the task in terms of matching the learner's cognitive and motivational profile.}
\end{flushleft}
\end{table}

\section{Appendix D}
\label{sec:app_d}
Appendix D presents the agent feedback gained through 10 iterations.

\begin{center}
\captionsetup{labelfont={it}, textfont={it}}
\captionof{table}{Stability of output verified by agent evaluation by 10 iterations}
\label{tab:agent_iterations}
%\small
\renewcommand{\arraystretch}{1}
    \begin{tabular}{@{}c>{\centering\arraybackslash}p{0.8cm}>{\centering\arraybackslash}p{0.8cm}>{\centering\arraybackslash}p{0.8cm}>{\centering\arraybackslash}p{0.8cm}>{\centering\arraybackslash}p{0.8cm}@{}}
    \toprule
     WS &  Iter &  Didact. &  Clarity &  Creat. &  Suitab. \\
    \midrule
             1 &          1 &                   5.0 &                  6.0 &         5.0 &                                   6.0 \\ 
             1 &          2 &                   6.0 &                  6.0 &         5.0 &                                   6.0 \\
             1 &          3 &                   6.0 &                  6.0 &         5.0 &                                   6.0 \\
             1 &          4 &                   5.0 &                  6.0 &         5.0 &                                   6.0 \\
             1 &          5 &                   6.0 &                  6.0 &         5.0 &                                   6.0 \\
             1 &          6 &                   5.0 &                  6.0 &         4.0 &                                   6.0 \\
             1 &          7 &                   5.0 &                  6.0 &         5.0 &                                   6.0 \\
             1 &          8 &                   5.0 &                  6.0 &         4.0 &                                   6.0 \\
             1 &          9 &                   5.0 &                  6.0 &         5.0 &                                   5.0 \\
             1 &         10 &                   6.0 &                  6.0 &         5.0 &                                   6.0 \\ \midrule
             2 &          1 &                   5.0 &                  6.0 &         5.0 &                                   6.0 \\
             2 &          2 &                   5.0 &                  6.0 &         5.0 &                                   6.0 \\
             2 &          3 &                   5.0 &                  6.0 &         5.0 &                                   6.0 \\
             2 &          4 &                   5.0 &                  6.0 &         5.0 &                                   6.0 \\
             2 &          5 &                   5.0 &                  6.0 &         5.0 &                                   6.0 \\
             2 &          6 &                   6.0 &                  6.0 &         6.0 &                                   6.0 \\
             2 &          7 &                   5.0 &                  6.0 &         5.0 &                                   6.0 \\
             2 &          8 &                   5.0 &                  5.0 &         5.0 &                                   6.0 \\
             2 &          9 &                   5.0 &                  6.0 &         5.0 &                                   6.0 \\
             2 &         10 &                   5.0 &                  6.0 &         5.0 &                                   6.0 \\ \midrule
             3 &          1 &                   5.0 &                  5.0 &         5.0 &                                   6.0 \\
             3 &          2 &                   5.0 &                  6.0 &         5.0 &                                   6.0 \\
             3 &          3 &                   6.0 &                  6.0 &         5.0 &                                   6.0 \\
             3 &          4 &                   5.0 &                  6.0 &         5.0 &                                   6.0 \\
             3 &          5 &                   5.0 &                  6.0 &         5.0 &                                   6.0 \\
             3 &          6 &                   5.0 &                  6.0 &         5.0 &                                   6.0 \\
             3 &          7 &                   5.0 &                  6.0 &         4.0 &                                   6.0 \\
             3 &          8 &                   5.0 &                  6.0 &         5.0 &                                   6.0 \\
             3 &          9 &                   5.0 &                  6.0 &         5.0 &                                   6.0 \\
             3 &         10 &                   5.0 &                  6.0 &         5.0 &                                   6.0 \\ \midrule
             4 &          1 &                   5.0 &                  6.0 &         5.0 &                                   6.0 \\
             4 &          2 &                   5.0 &                  6.0 &         5.0 &                                   6.0 \\
             4 &          3 &                   5.0 &                  6.0 &         5.0 &                                   6.0 \\
             4 &          4 &                   5.0 &                  6.0 &         4.0 &                                   6.0 \\
             4 &          5 &                   5.0 &                  6.0 &         5.0 &                                   6.0 \\
             4 &          6 &                   5.0 &                  6.0 &         5.0 &                                   6.0 \\
             4 &          7 &                   5.0 &                  6.0 &         5.0 &                                   6.0 \\
             4 &          8 &                   5.0 &                  6.0 &         5.0 &                                   6.0 \\
             4 &          9 &                   5.0 &                  6.0 &         5.0 &                                   6.0 \\
             4 &         10 &                   5.0 &                  6.0 &         5.0 &                                   6.0 \\
\bottomrule
\end{tabular}
\begin{flushleft}
{\footnotesize 
\textit{Note}: Internal validity was evaluated by performing 10 iterations per worksheet. WS = worksheet, Iter = iteration, Didact. = didactical structure, Clarity = clarity of the task, Creat. = creativity, Suitab. = suitability for learners. Evaluation based on a 6-point ranking scale (1 = insufficient, 6 = very good). Evaluation of the worksheets for the four example learner groups: Learner 1 = Limited Knowledge with High Motivation, Learner 2 = Limited Knowledge with Low Motivation, Learner 3 = Advanced Knowledge with Low Motivation, Learner 4 = Advanced Knowledge with High Motivation.}
\end{flushleft}
\end{center}

\end{document}